\documentclass[a4paper]{article}

\usepackage{INTERSPEECH2020}

\usepackage{color}
\usepackage{diagbox}
\usepackage{xcolor,colortbl}
\usepackage{enumitem}
\usepackage{amssymb}
\usepackage{balance}

\usepackage{times}
\usepackage{graphicx}
\usepackage{amssymb}
\usepackage{gensymb}
\usepackage{amsmath}
\usepackage{caption}
\usepackage{subcaption}
\usepackage{tikz}
\usetikzlibrary{automata, positioning, arrows}

\usepackage{url,hyperref}

\usepackage{algorithm}
\usepackage{algorithmic}

\usepackage[labelfont={bf},font=small]{caption}
\usepackage[none]{hyphenat}

\usepackage{mathtools, cuted}

\usepackage[noadjust, nobreak]{cite}

\usepackage{tabularx}
\usepackage{amsmath}

\usepackage{float}

\usepackage{pifont}

\newcolumntype{Y}{>{\centering\arraybackslash}X}

\usepackage[]{placeins}

\usepackage{placeins}

\usepackage{tikz}

\usepackage[framemethod=tikz]{mdframed}

\usepackage{afterpage}

\usepackage{stfloats}

\usepackage{atbegshi}
\newcommand{\handlethispage}{}
\newcommand{\discardpagesfromhere}{\let\handlethispage\AtBeginShipoutDiscard}
\newcommand{\keeppagesfromhere}{\let\handlethispage\relax}
\AtBeginShipout{\handlethispage}

\usepackage{comment}

\title{``I have vxxx bxx connexxxn!'':\\ Facing Packet Loss in  Deep Speech Emotion Recognition}

\name{Mostafa M.\ Mohamed$^{1,2}$ and Bj\"orn W.\ Schuller$^{2,3}$}
\address{
  $^1$Chair of Embedded Intelligence for Health Care and Wellbeing, University of Augsburg, Germany\\
  $^2$AI R\&D Team, SyncPilot GmbH, Augsburg, Germany\\
  $^3$GLAM -- Group on Language, Audio, \& Music, Imperial College London, UK}
\email{mostafa.amin@syncpilot.com, schuller@IEEE.org}

\begin{document}

\maketitle
\thispagestyle{empty}

\begin{abstract}
In applications that use emotion recognition via speech, frame-loss can be a severe issue given manifold applications, where the audio stream loses some data frames, for a variety of reasons like low bandwidth. In this contribution, we investigate for the first time the effects of frame-loss on the performance of emotion recognition via speech. Reproducible extensive experiments are reported on the popular RECOLA corpus using a state-of-the-art end-to-end deep neural network, which mainly consists of convolution blocks and recurrent layers. A simple environment based on a Markov Chain model is used to model the loss mechanism based on two main parameters. We explore matched, mismatched, and multi-condition training settings. As one expects, the matched setting yields the best performance, while the mismatched yields the lowest. Furthermore, frame-loss as a data augmentation technique is introduced as a general-purpose strategy to overcome the effects of frame-loss. It can be used during training, and we observed it to produce models that are more robust against frame-loss in run-time environments.
\end{abstract}


\section{Introduction}
There is a rise of affective computing applications which predict emotions through speech or other signals like images. Such applications depend heavily on the quality of the audio streams of speech to correctly predict the emotions. 
In streaming applications, there are a variety of factors that could result in lower quality of data received, like lower data rate and packet loss \cite{buffering}, or varying throughput in mobile communication \cite{schmid2019deep}.
In such applications, any issue like this that might happen, would cause a drop in the input streams which might lead to severe degradation in the performance of the application.
Such a degradation could happen for a variety of reasons, for example, the dependency of some models on the audio context to predict the emotions of the succeeding time points, also when some models assume the continuity of the input speech. These are typical assumptions made by neural network models like \cite{tzirakis2018,  Adieu}, because of the design of recurrent neural networks \cite{deeplearn}.

The impact of disturbances during automatic `speech emotion recognition' (SER) has been investigated for speech in the presence of noise \cite{Schuller06-ERI,Weninger11-RON,Pohjalainen16-SAC}, reverberation \cite{Schuller11-ASS,Weninger11-RON}, or in narrowband transmission \cite{Marchi16-TEO} and coded speech \cite{albin2015objective,Marchi16-TEO}. However, to the authors' best knowledge, no work exists that investigates the impact of packet (or frame) loss on SER. There are only a few papers addressing SER in VoIP setting \cite{pao2012integration}, yet, not systematically investigating packet loss impact. This seems surprising, given that a main application of SER is found in call centres, and SER is currently finding its way onto mobile phones \cite{Marchi16-RTO}.  Packet loss and its impact on speech processing has largely been studied so far in the context of automatic speech recognition \cite{milner2001robust} and enhanced in \cite{Lotfidereshgi2018}.


The main aim of this paper is to examine the effects of these frame-loss cases on the performance of models for emotion recognition via speech. In addition to that, an attempt to enhance such models to become more robust against frame-loss will be made.

This paper is divided as follows: Section \ref{sec:approach} contains the details of the approach, Section \ref{sec:results} contains the experimental settings and the results, and Section \ref{sec:conclusion} provides the conclusion of the paper.


\section{Approach}
\label{sec:approach}
The approach mainly uses an end-to-end model which predicts emotions (defined as two dimensions {\it arousal} and {\it valence}). The model is trained and tested under a variety of settings that are simulated by a mechanism modelling lossy environments.


\begin{figure}[t!]
    \centering
    \includegraphics[width=\linewidth]{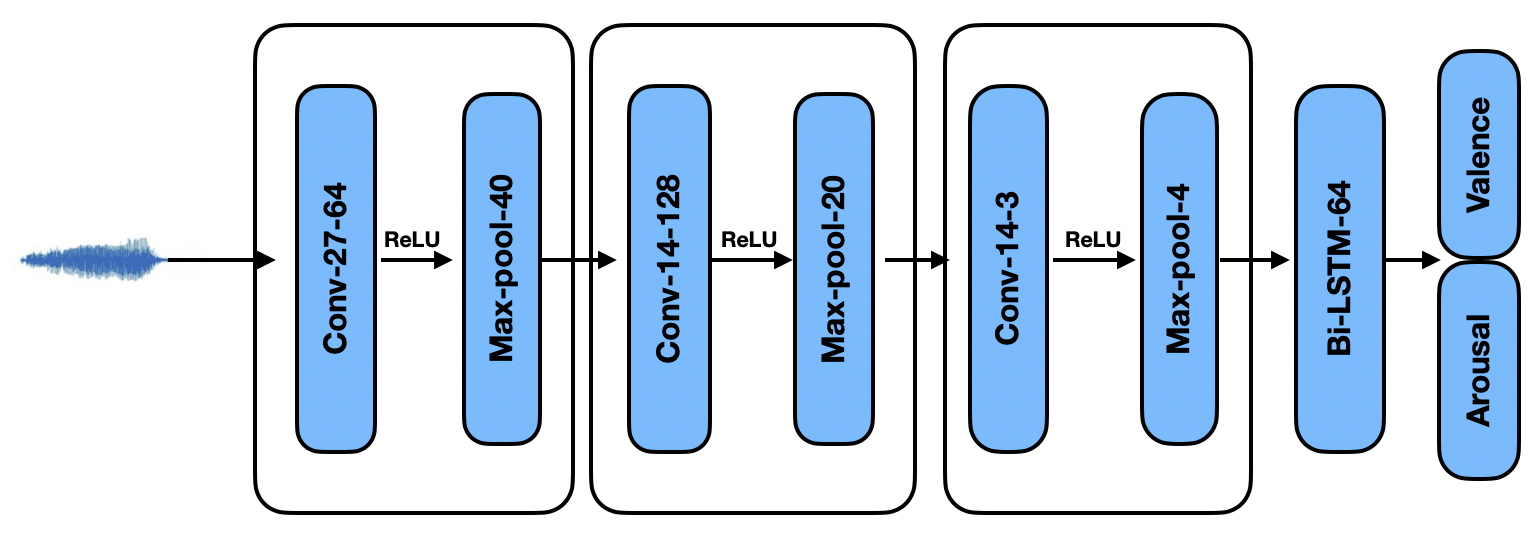}
    \caption{End-to-end model for speech emotion recognition.}
    \vspace{-0.25cm}
    \label{fig:model}
\end{figure}

\subsection{Packet loss generation model}

\begin{figure}[t!]
    \centering
    \begin{tikzpicture}
        \node[state, initial] (qn) {$N:1$};
        \node[state, right of=q2] at (2, 0) (ql) {$L:0$};
        
        \draw 
        (qn) edge[loop above] node{$p_{\text{N}}$} (qn)
        (ql)[->] edge[loop above] node{$p_{\text{L}}$} (ql)
        (qn)[->] edge[bend left, above] node{$1 - p_{\text{N}}$} (ql)
        (ql) edge[bend left, below] node{$1 - p_{\text{L}}$} (qn);
    \end{tikzpicture}

    \caption{Markov Chain $\mathcal{M}(p_{\text{L}}, p_{\text{N}})$ that samples a binary sequence,  that can be used as a mask for {\it loss} or {\it non-loss} combinations.}
        \label{fig:markov}
\end{figure}
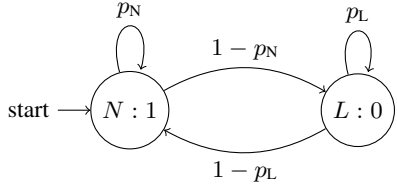

In order to model the lossy and non-lossy packets -- or, more precisely, frames in our case -- in a given sequence, we adapt the Markov Chain \cite{bishop2006pattern} $\mathcal{M}(p_{\text{L}}, p_{\text{N}})$ as shown in Figure \ref{fig:markov}. This is a standard approach for packet loss modelling \cite{gilbert}; note, however, that also more complex models, e.\,g., three states have been used \cite{milner2004analysis}, for example, to model burst behaviour. Other models are also reviewed in a recent survey \cite{da2019mac}.
Given a sequence of $t$ frames, we can use $\mathcal{M}$ to sample a binary sequence of length $t$. This can be achieved by starting at the state $N$, then transitioning between the states $N$ (for no-loss) and $L$ (for loss) based on the transition probabilities $p_{\text{L}}$ and $ p_{\text{N}}$. This is done until $t$ states are enumerated. Then, the sampled sequence of states is directly transformed into the binary string, by replacing $N$ by $1$ and $L$ by $0$.

The sampled binary string can be used to select elements from the given sequence, where the frames at positions with corresponding character `1' are the only frames taken. For example, a binary sequence $01101$ would select the frames $y_2, y_3, y_5$ from the sequence $y_1,y_2,y_3,y_4,y_5$.

The intuition behind this model is that it can mimic a variety of possibilities. The value of $p_{\text{N}}$ models the overall stability of the system, in particular, how unlikely it is that a frame-loss error might occur. Additionally, the value of $p_{\text{L}}$ models the intensity of frame-loss when it occurs. High values of $p_{\text{L}}$ correspond to persistent errors that stay long. Different combinations of these can correspond to different possibilities as shown in Table \ref{tab:markov}. An environment with a low bandwidth could be thought of as to have low values for both parameters, which mirrors a scenario of frequent non-persistent frame-loss issues. If both parameters have high values, this mirrors an environment with a low chance of a persistent breakdown event.

\begin{table}[t!]
    \centering
    \begin{tabular}{l|l|l}
         & low $p_{\text{L}}$ & high $p_{\text{L}}$ \\
        \hline
        
    high $p_{\text{N}}$ & stable & sudden breakdown \\
    low $p_{\text{N}}$ & low bandwidth & extremely unstable \\
    
    \end{tabular}
    \vspace{.25cm}
    \caption{How different values of $p_{\text{N}}$ and $p_{\text{L}}$ may model different environments with different causes for frame-loss.}
    \vspace{-0.5cm}
    \label{tab:markov}
\end{table}

Furthermore, we will need to drop frames from two sequences simultaneously, mainly when one is an input audio sequence $X$ and the other is $Y$ which consists of the output labels sequence. Even though, both correspond to the same duration of time, still the sample rate of $X$ is higher than that of $Y$. For simplicity, it is assumed that the sample rate of $X$ is a multiple of the sample rate of $Y$, with a multiplicative factor $r$. Based on that assumption, if we acquired a binary string $M_Y$ from the model $\mathcal{M}$ to drop the frames of the output labels $Y$, then we can construct a mask $M_X$ to drop the corresponding elements of $X$. The mask $M_X$ is constructed by repeating each element of $M_Y$ for $r$ times in place. For example, if $r = 3$ and the mask `1011' is used to drop frames from $Y$, then $X$ is dropped using `111000111111'. This mechanism ensures that the dropping of frames corresponds to the same time tags. Eventually, the given Markov Chain $\mathcal{M}(p_L, p_N)$ will sample binary strings that have an expected fraction of losses \cite{xiao2018packet}: 
\begin{equation}
    \frac{1 - p_N}{2 - p_L - p_N}.
\end{equation}

\subsection{Dataset}
The dataset that is used in the experiment is the RECOLA dataset \cite{RECOLA}. The training data consists of the 16 training tracks, 15 validation tracks, and 15 test tracks. Each track consists of 5 minutes of audio \cite{RECOLA}, recorded at 44.1\,kHz. Each track is labelled across time and the labels were collected at a frequency of 25\,Hz.
Each track contains one student participant with a mean age of 22 years. The speakers spoke in a variety of languages which consisted of 33 French, 8 Italian, 4 German, and 1 Portuguese speakers. In our experiments, the audio tracks are down-sampled to 16\,kHz, and the labels are down-sampled by a factor of $5$ using median pooling.
Since the labels for the test portion were not freely accessible at the time of the experiments, the validation portion is used for testing.
\subsection{Model}

There needs to be a model that can recognise emotions via speech, where emotions are defined by two main dimensions {\it arousal} and {\it valence}. For this purpose, an end-to-end deep model is used, due to its simplicity and strong performance. There is one model architecture that is adapted in all the experiments, based on a variant of the model introduced by \cite{tzirakis2018}, with slightly different hyperparameters.

The model's architecture is depicted in Figure \ref{fig:model}. It starts with a batch normalisation layer \cite{batchnorm}, followed by three convolution blocks, then a bidirectional LSTM layer \cite{LSTM}, and finally a time-distributed fully-connected layer \cite{FC} (using $\tanh$ activation function) with two output features. Bidirectional LSTMs have shown to be effective in ASR \cite{bilstm}. Each of the convolution blocks or the recurrent layers are followed by a dropout layer (dropout rate $0.5$) to reduce overfitting \cite{dropout}. Each convolution block consists of a 1D convolution layer (with $ReLU$ activation function) followed by a max-pooling layer. The convolution layers have filter sizes 27, 14, and 3 respectively. The number of output channels are 64, 128, and 128 respectively. The pooling sizes are 40, 20, and 4 respectively. The bidirectional LSTM consists of 64 output units.
The sizes of the pooling layers are chosen to reduce the input sample rate from 16\,kHz to an output sample rate of 5\,Hz. Accordingly, the kernel layers have a padding to preserve the input length. Then, their filters' sizes are chosen to render the overlap rate $R \approx 0.4$ as advised in \cite{tzirakis2018}. The overlap rate is calculated by the formula:
\begin{equation}
    R = \frac{K - 1}{K + P - 1}.
\end{equation}
\label{sec:training}
During training, the input and output data are segmented into frames of 20 seconds, in order to reduce the time complexity needed by the LSTM layers to operate on long sequences. The training is performed using the Adadelta optimisation algorithm \cite{adadelta} with a learning rate of $0.5$, for $200$ epochs and a mini-batch size of $16$. Similar to \cite{tzirakis2018}, the loss function that is used for training is a function that would maximise the {\it concordance correlation coefficient} (CCC) \cite{ccc}. The function is $1 - \rho_c(y, \hat{y})$, where $\rho_c$ is the CCC, defined by the formula:
\begin{equation}
\rho_c(x, y) = \frac{2\sigma_{xy}^2}{\sigma_x^2 + \sigma_y^2 + (\mu_x - \mu_y)^2},
\label{eq:CCC}
\end{equation}
where $\sigma_x^2, \sigma_y^2$ are the variances of $x$ and $y$ respectively, 
$\mu_x, \mu_y$ are the means of $x$ and $y$ respectively, and $\sigma_{xy}^2$ is the covariance of $x$ and $y$. The loss function uses the CCC on the time dimension of the data, then averages the values across examples and emotions features, in order to ensure that both emotion dimensions are optimised adequately.

\section{Experiments and Results}
\label{sec:results}

\begin{figure*}[t!]
  \centering
  \includegraphics[width=.8\textwidth]{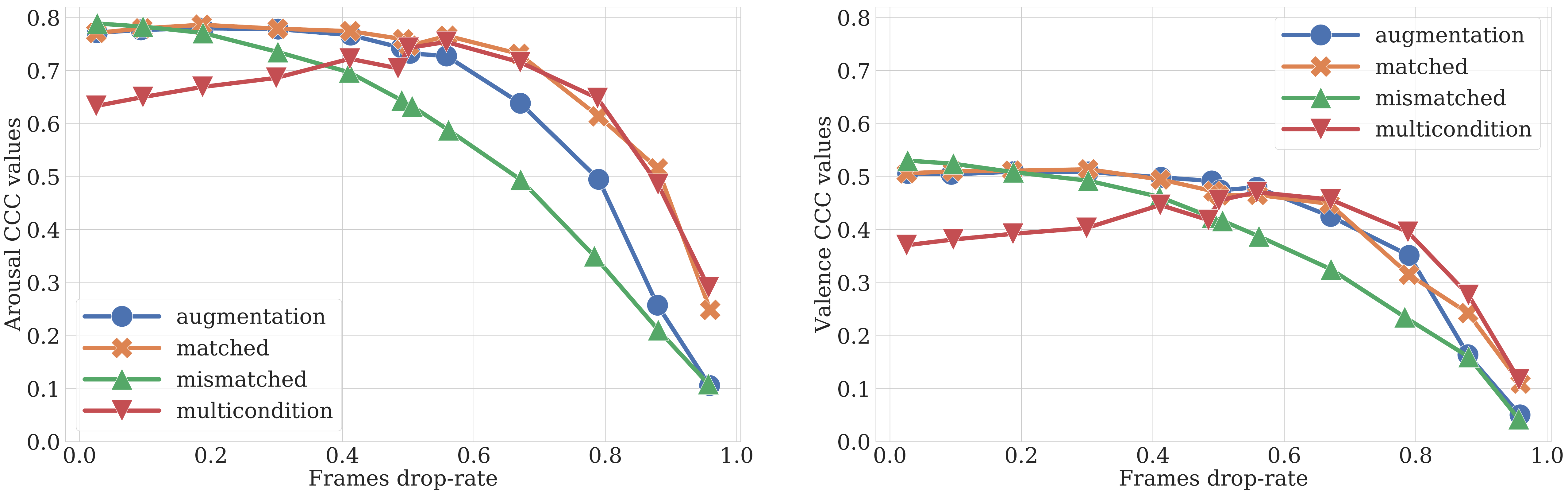}
 \caption{CCC scores for arousal and valence compared against different frame-drop rates, for the different training settings.}

\label{fig:ccc_line}
\end{figure*}

\subsection{Experimental settings}

The effects of frame-loss on emotion recognition are investigated under four different settings: matched, mismatched, multi-conditions, and augmentation. The main difference between these settings is the training environment. Table \ref{tab:training} shows the validation CCC scores for all the chosen settings. The testing environment is the same for all of them; it considers several combinations of the two parameters $p_{\text{N}}$ and $ p_{\text{L}}$. Depending on the chosen values for both parameters and the training setting, a corresponding model is chosen to be tested using CCC (in Equation \ref{eq:CCC}). The testing is done by applying the frame-loss (in the corresponding settings only) individually on each of the five minutes tracks, then predicting the labels for the remaining frames. The comparison between labels and predictions is then done individually for each emotion dimension, by calculating CCC on the concatenation of all tracks (since they might have different lengths after applying frame-loss).

\subsubsection{Mismatched training}
In the mismatched setting, the training is run on the clean data without any application of frame-loss, and the same model is used for all test combinations.

\subsubsection{Multi-conditions training}
In the multi-conditions training settings, for each training batch, two values $p_{\text{N}}$ and $p_{\text{L}}$ are sampled uniformly from $[0.05, 1]$ and $[0, 1]$ respectively. Then, accordingly, a frame-loss mask is sampled using the Markov Chain $\mathcal{M}$. The sampled mask is used to drop frames for all the examples in the batch. Only one model is trained in this setting, and it is used for all test combinations. During sampling, $p_{\text{N}}$ is clipped to be at least $0.05$ to prevent extremely high loss of training data which degrades the training quality severely.

\subsubsection{Matched training}
The training environment in the matched settings relies on partial multi-conditions training, because there are many test combinations of the two parameters $p_{\text{N}} $ and $p_{\text{L}}$, and it would be impractical to train a model for each of those combinations. Consequently, the values are clustered in three categories: {\it low}, {\it medium}, and {\it high}, with values in the ranges $[0, 1/3)$, $[1/3, 2/3)$ and $[2/3, 1]$, respectively. Using these categories, there are nine combinations for models to be trained. In each combination, based on the chosen categories, values for both $p_{\text{N}}$ and $p_{\text{L}}$ are sampled uniformly for each batch (according to the corresponding categories' ranges). Similar to the multi-conditions setting, according to sampled values of $p_{\text{N}}$ and $p_{\text{L}}$, a mask is generated using the introduced Markov Chain $\mathcal{M}$ to drop the frames of the whole batch. $p_{\text{N}}$ is again clipped to be at least $0.05$ to prevent the severe degradation of training quality. However, still some residues of the degradation is visible in the last row of Table \ref{tab:training}. During the testing, depending on the categories in which each of the testing values of $p_{\text{N}}$ and $p_{\text{L}}$ lie in, the model with the corresponding matching category is chosen for testing.

\subsubsection{Augmentation training}
In this setting, one of the models from the matched training setting is used, when $p_{\text{L}}$ is low and $p_{\text{N}}$ is high. This one model is then used for all the test combinations. This setup is similar to the multi-conditions setup, with one key difference, which is the model used for testing. The main aim of this setting is to examine the effectiveness of a frame-loss as a data augmentation technique \cite{augmentation} which can be used during training with the aim to improve the results or allow the model to be more robust in degraded run-time environments.

\begin{table}[t!]
\centering
\begin{tabular}{l|l l|r r}

setting & $p_{\text{N}}$ & $p_{\text{L}}$ & arousal & valence \\
\hline
mis & - & - & .789 & {\bf .529} \\ 
\cite{tzirakis2018} & - & - & {\bf .815} & .502 \\
multi & $[0.05, 1]$ & $[0, 1]$ & .630 & .366 \\
match & {\it high} & {\it mid} & .797 & .542 \\
match/aug & {\it high} & {\it low} & .769 & .503 \\
match & {\it mid} & {\it low} & .736 & .501 \\
match & {\it high} & {\it high} & .729 & .489 \\
match & {\it mid} & {\it high} & .702 & .452\\
match & {\it mid} & {\it mid} & .701 & .425 \\
match & {\it low} & {\it low} & .662 & .426 \\
match & {\it low} & {\it mid} & .650 & .405 \\
match & {\it low} & {\it high} & .430 & .176 \\


\end{tabular}
\vspace{.15cm}
\caption{CCC scores on validation data (without any frame-loss) for the different training settings. The values {\it mid} and {\it high} correspond to the ranges
$[1/3, 2/3)$ and $[2/3, 1]$ respectively, while {\it low} corresponds to the range $[0, 1/3)$ for $p_{\text{L}}$, and $[0.05, 1/3)$ for $p_{\text{N}}$. \cite{tzirakis2018} is shown in the second row.}
\vspace{-0.75cm}
\label{tab:training}
\end{table}

\subsection{Results}

\begin{figure*}[t!]
  \centering
  \includegraphics[width=0.8\textwidth]{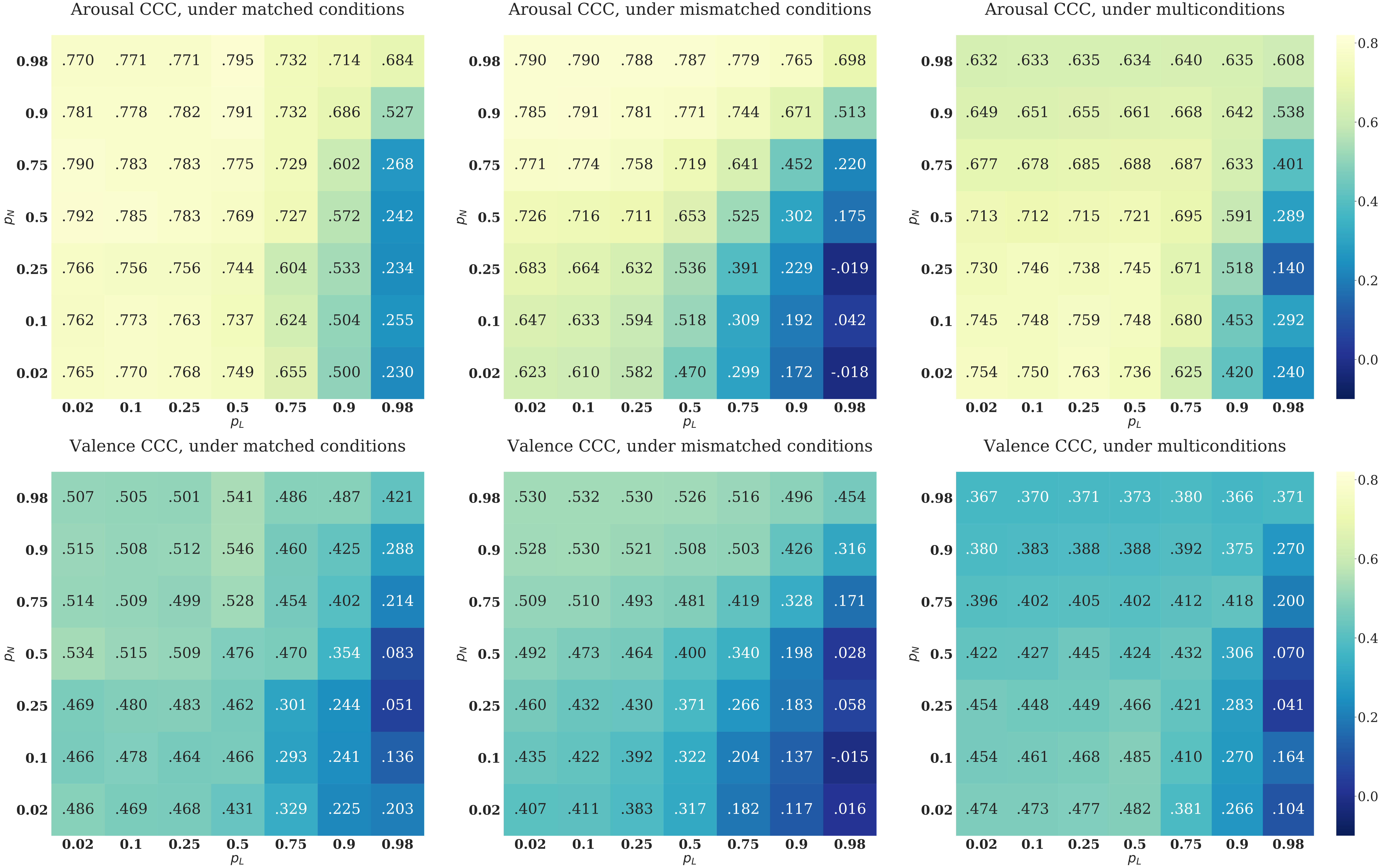}
  \caption{CCC scores for valence and arousal, for the three matched, mismatched, and multi-conditions settings. $p_{\text{N}}$ is the probability to remain in a {\it non-loss} state, $p_{\text{L}}$ is the probability of remain in a {\it loss} state.}
  \label{fig:ccc_matrix}
\end{figure*}

 The results in Figure \ref{fig:ccc_line} are comparing the scores to a single dimension, which is the ratio of dropped frames after applying the frame-loss. The results of the testing are shown in details in Figure \ref{fig:ccc_matrix}, where the different combinations of valence/arousal and the three training settings (matched, mismatched, and multi-conditions) are examined.

According to Figure \ref{fig:ccc_line}, it can be seen that generally, the matched setting has the overall best performance, while the mismatched has the worst overall performance. The performance of the matched setting is expected since the model gets trained on data which is the most similar to the test data, in comparison to the other settings. In addition, for a low drop-rate $<0.5$, the multi-conditions setting tends to have the worst performance, while the matched and mismatched settings are more or less on par.

The previous results were the main motivation to examine the augmentation setting, which tries to combine the advantages of the mismatched settings and multi-conditions, without matching the training and testing. In that case, one model is trained with parameters that cause a low drop-rate. The aim is to achieve the high performance of the matched settings for the low drop-rate, and resembles some of the high performance of the multi-conditions setting on the high drop-rate. The results according to Figure \ref{fig:ccc_line} show that this is indeed the case. The augmentation setting achieves nearly similar performance like the matched setting for drop-rate $<0.5$, while making some improvement over the mismatched setting for higher drop-rate.

After examining the results of the different settings, a strategy to overcome the frame-loss effects is to try to match the setting of the training environment to match the deployment environment. In case this matching is hard to be performed, a data augmentation technique can be a general purpose technique to use. For particular environments with severe degradation in the audio's quality, the training with multi-conditions setting can then be used.

\section{Conclusions}
\label{sec:conclusion}
In this paper, the effects of frame-loss on the performance of automatic speech emotion recognition were examined. A Markov Chain model was utilised to model environments with frame-loss, where an audio stream can lose data packets during transmission. For such an examination, an end-to-end deep model was used for the experiments. The model mainly consists of convolution blocks and recurrent layers and the dataset RECOLA was chosen for the experiments.

The experiments had mainly three settings: matched, mismatched, and multi-conditions settings. In all of the settings, the models were tested with a variety of possibilities of frame-loss, while the training was the crucial difference between the different settings. In the mismatched setting, the model was trained on clean data. In the matched setting, a variety of models were trained based on {\it low}, {\it mid}, or {\it high} values of the parameters. In the multi-conditions settings, one model was trained using a mixture of all parameters' combinations.

The results have shown that the matched settings had the best overall performance while the mismatched setting had the worst overall performance. The multi-conditions setting was on par with the matched settings for lossy data (with frame-loss rate $>0.5$). However, it was the worst on data with low frame-loss rate $<0.5$. On the other hand, the matched and mismatched settings had an on par performance for data with low frame-loss rates $<0.5$.

An additional setting was experimented to test out a general purpose solution for the frame-loss problem, namely training with frame-loss as a data augmentation mechanism, just using parameters that lead to low frame-loss rates. The augmentation has been shown as a compromise strategy to combine the advantages of the mismatched and multi-conditions settings, without matching the training to the test environments. It has shown a performance on low frame-loss rates which is on par to the matched setting, while for high frame-loss rates it has shown an improvement over the mismatched setting.  

Future work should investigate the use of Packet Loss Concealment (PLC) methods \cite{rodbro2006hidden} in the context of SER instead of classical PLC \cite{hmminterspeech}. This could include recent deep learning approaches including such from the image processing domain \cite{athar2018latent} originally tailored for occlusion restoration, as it has repeatedly been shown that audio can well be modelled as an `image' using the spectogram or related representations \cite{Cummins17-AID}.

\bibliographystyle{IEEEtran}
\bibliography{ms}




\clearpage

\end{document}